\documentclass[twocolumn,prl,superscriptaddress,amsmath,amssymb]{revtex4}

\usepackage{graphicx}
\usepackage{dcolumn}
\usepackage{bm}
\usepackage{epstopdf}

\begin{document}

\title{Pressure-driven collapse of the relativistic electronic ground state \\in a honeycomb iridate}

\author{J. P. Clancy}
\affiliation{Department of Physics, University of Toronto, Toronto, Ontario M5S~1A7, Canada}
\author{H. Gretarsson}
\affiliation{Department of Physics, University of Toronto, Toronto, Ontario M5S~1A7, Canada}
\author{J. A. Sears}
\affiliation{Department of Physics, University of Toronto, Toronto, Ontario M5S~1A7, Canada}
\author{Yogesh Singh}
\affiliation{Indian Institute of Science Education and Research Mohali, Sector 81, SAS Nagar, Manauli PO 140306, India}
\author{S. Desgreniers}
\affiliation{Laboratoire de physique des solides denses, Department of Physics, University of Ottawa, Ottawa, Ontario, K1N 6N5, Canada}
\author{Kavita Mehlawat}
\affiliation{Indian Institute of Science Education and Research Mohali, Sector 81, SAS Nagar, Manauli PO 140306, India}
\author{Samar Layek}
\affiliation{School of Physics and Astronomy, Tel Aviv University, 69978 Tel Aviv, Israel}
\author{Gregory Kh. Rozenberg}
\affiliation{School of Physics and Astronomy, Tel Aviv University, 69978 Tel Aviv, Israel}
\author{Yang Ding}
\affiliation{Center for High-Pressure Science \& Technology Advanced Research (HPSTAR), Beijing, 100094, China}
\author{M. H. Upton}
\affiliation{X-ray Science Division, Advanced Photon Source,
Argonne Nation Laboratory, Argonne, Illinois 60439, USA}
\author{D. Casa}
\affiliation{X-ray Science Division, Advanced Photon Source,
Argonne Nation Laboratory, Argonne, Illinois 60439, USA}
\author{N. Chen}
\affiliation{Canadian Light Source, Saskatoon, Saskatchewan, S7N 0X4, Canada}
\author{Junhyuck Im}
\affiliation{Department of Earth System Sciences, Yonsei University, Seoul 120-749, Korea}
\author{Yongjae Lee}
\affiliation{Department of Earth System Sciences, Yonsei University, Seoul 120-749, Korea}
\affiliation{Center for High Pressure Science \& Technology Advanced Research (HPSTAR), Shanghai 201203, China}
\author{R. Yadav}
\affiliation{Institute for Theoretical Solid State Physics, IFW Dresden, Helmhotzstr. 20, 01069 Dresden, Germany}
\author{L. Hozoi}
\affiliation{Institute for Theoretical Solid State Physics, IFW Dresden, Helmhotzstr. 20, 01069 Dresden, Germany}
\author{D. Efremov}
\affiliation{Institute for Theoretical Solid State Physics, IFW Dresden, Helmhotzstr. 20, 01069 Dresden, Germany}
\author{J. van den Brink}
\affiliation{Institute for Theoretical Solid State Physics, IFW Dresden, Helmhotzstr. 20, 01069 Dresden, Germany}
\author{Young-June Kim}
\email{yjkim@physics.utoronto.ca} \affiliation{Department of Physics, University of Toronto, Toronto, Ontario M5S~1A7, Canada}

\date{\today}

\begin{abstract}
The electronic ground state in many iridate materials is described by a complex wave-function in which spin and orbital angular momenta are entangled due to relativistic spin-orbit coupling (SOC) \cite{Witczak2014, Rau2016}. Such a localized electronic state carries an effective total angular momentum of $J_{eff}=1/2$ \cite{BJKim2008, BJKim2009}.  In materials with an edge-sharing octahedral crystal structure, such as the honeycomb iridates $\rm Li_2IrO_3$ and $\rm Na_2IrO_3$, these $J_{eff}=1/2$ moments are expected to be coupled through a special bond-dependent magnetic interaction \cite{Jackeli2009, Chaloupka2010, Singh2012}, which is a necessary condition for the realization of a Kitaev quantum spin liquid \cite{Kitaev2006}.  However, this relativistic electron picture is challenged by an alternate description, in which itinerant electrons are confined to a benzene-like hexagon, keeping the system insulating despite the delocalized nature of the electrons \cite{Mazin2012, Mazin2013}.  In this quasi-molecular orbital (QMO) picture, the honeycomb iridates are an unlikely choice for a Kitaev spin liquid.  Here we show that the honeycomb iridate $\rm Li_2IrO_3$ is best described by a $J_{eff}=1/2$ state at ambient pressure, but crosses over into a QMO state under the application of small ($\sim$0.1~GPa) hydrostatic pressure.  This result illustrates that the physics of iridates is extremely rich due to a delicate balance between electronic bandwidth, spin-orbit coupling, crystal field, and electron correlation.
\end{abstract}

\maketitle

Understanding the physical properties of a condensed matter system is greatly influenced by the choice of basis used to describe its electronic state: a localized wavefunction or an itinerant one.  This distinction is not always straightforward due to the complex hierarchy of energy scales involved, and phase sensitive experimental methods are not always available to probe the wavefunction of a given system directly.  However, the nature of the wavefunction can often be revealed indirectly when an appropriate tuning parameter can be used to vary the electronic properties of the system.  Hydrostatic pressure is a particularly effective tuning parameter, as it can be used to directly modify the overlap between electronic orbitals, and thereby control the electronic bandwidth \cite{Haskel2012, Tafti2012, Clancy2016}.

We have studied the evolution of the structural and electronic properties of the honeycomb lattice iridate $\alpha$-$\rm Li_2IrO_3$ as a function of applied hydrostatic pressure using three complementary synchrotron x-ray techniques.  We used conventional x-ray powder diffraction (XRD) to study the crystal structure of this material, resonant inelastic x-ray scattering (RIXS) to investigate the electronic excitation spectrum, and x-ray absorption spectroscopy (XAS) to probe the strength of the SOC.  The pressure of the sample was tuned from ambient pressure up to 10~GPa using a diamond anvil cell (DAC).

\begin{figure*}
\begin{center}
\includegraphics[angle=0,width=\textwidth]{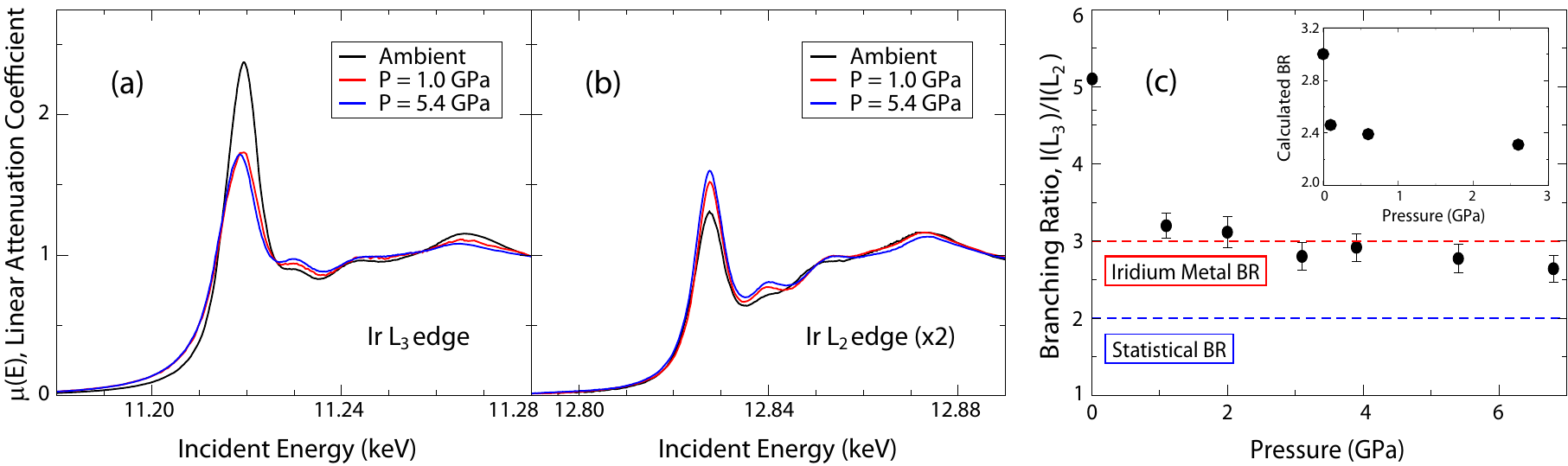}
\end{center}
\caption{X-ray absorption spectroscopy (XAS) can be used to probe the relativistic $J_{eff}=1/2$ ground state of $\alpha$-$\rm Li_2IrO_3$.  At ambient pressure, XAS measurements reveal an anomalously large intensity difference between the main white line features observed at (a) the Ir L$_3$ ($2p_{3/2}$ $\rightarrow$ $5d$) and (b) Ir L$_2$ ($2p_{1/2}$ $\rightarrow$ $5d$) absorption edges.  This large L$_3$/L$_2$ branching ratio is a strong signature of the $J_{eff}=1/2$ ground state.  (c) Under applied pressure, the branching ratio decreases rapidly, approaching a value reminiscent of elemental iridium.  The high pressure branching ratio remains greater than the statistical ratio of 2, but falls well below the values reported for other spin-orbit-driven $J_{eff}=1/2$ systems.  (Inset) This qualitative trend is also captured by quantum chemistry calculations based on the experimental crystal structure.}
\label{fig:xas}
\end{figure*}

One of the defining signatures of the $J_{eff}=1/2$ relativistic electronic ground state is an unusually large difference between the XAS “white line” intensity observed at the Ir L$_3$ ($2p_{3/2}$ $\rightarrow$ $5d$) and L$_2$ ($2p_{1/2}$ $\rightarrow$ $5d$) absorption edges \cite{Laguna2010, Haskel2012, Clancy2012}.  The ratio of these intensities is known as the L$_3$/L$_2$ branching ratio (BR), and provides a direct measure of $\langle {\bf L} \cdot {\bf S} \rangle$, the angular part of the expectation value for the spin orbit operator \cite{VanderLaan1988, Thole1988}.  The pressure dependence of the x-ray absorption spectra for $\alpha$-$\rm Li_2IrO_3$ is provided in Fig.~\ref{fig:xas}.  The large branching ratio observed at ambient pressure (BR = $5.1 \pm 0.4$) is consistent with a $J_{eff}=1/2$ state, and is similar to previously reported BR for other spin-orbit-driven iridates such as $\rm Sr_2IrO_4$ \cite{Haskel2012, Clancy2012}.  However, the BR of $\alpha$-$\rm Li_2IrO_3$ drops precipitously under applied pressure, falling to less than $2/3$ of its original value by P = 1.1 GPa.  The BR continues to decrease more gradually up to $\sim 3$~GPa, and ultimately plateaus at a high pressure value of $2.8 \pm 0.1$.  Although dramatically reduced from ambient pressure, this value still exceeds the statistical branching ratio (BR = 2) expected in the limit of negligible SOC.  In fact, it is strikingly similar to that of iridium metal (BR $\sim 3$) \cite{Clancy2012, Qi1987, Jeon1989}, a material which exhibits significant SOC, but which does not harbour a $J_{eff}=1/2$ ground state.  As a result, the XAS data suggests that applied pressure results in a collapse of the $J_{eff}=1/2$ ground state in $\alpha$-$\rm Li_2IrO_3$ by P = 1.1 GPa.

The abrupt drop in branching ratio is also qualitatively reproduced by {\it ab initio} quantum chemistry calculations, as shown in the inset of Fig. 1(c) and described in the Supplemental Material (SM).  In fact, these calculations, which are based on the experimental crystal structures determined from XRD, suggest that the drop in branching ratio actually occurs at significantly lower pressures, close to P = 0.1 GPa.  We have carried out high pressure electrical resistance measurements on $\alpha$-$\rm Li_2IrO_3$ (provided in the SM), which indicate that the sample remains insulating up to 7 GPa.  This confirms that it is the $J_{eff}=1/2$ character of the ground state, and not its insulating properties, that is disrupted by applied pressure.

The pressure scale associated with this change in BR is quite remarkable in comparison with other iridates.  In $\rm Sr_2IrO_4$ for example, the BR remains essentially unchanged up to 30 GPa \cite{Haskel2012}, and an applied pressure of 70 GPa is required to produce a decrease similar to what is observed in Fig.~\ref{fig:xas}.  This suggests that $\alpha$-$\rm Li_2IrO_3$ is situated much closer to the boundary of the $J_{eff}=1/2$ relativistic electronic state, and that it is possible to tune the system into a new electronic ground state under the influence of applied pressure.

\begin{figure*}
\begin{center}
\includegraphics[angle=0,width=\textwidth]{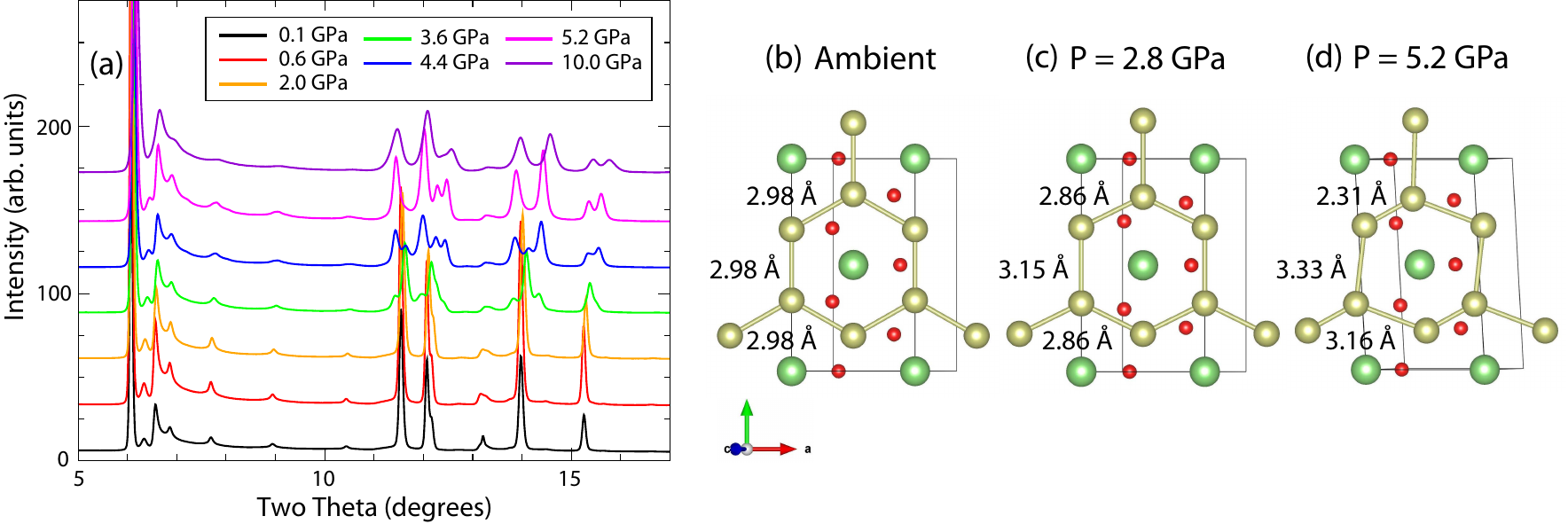}
\end{center}
\caption{ (a) High pressure x-ray powder diffraction measurements reveal a series of pressure-induced structural distortions in $\alpha$-$\rm Li_2IrO_3$.  (b) At ambient pressure, $\alpha$-$\rm Li_2IrO_3$ displays an almost ideal undistorted Ir honeycomb lattice.  (c) As the pressure increases, the honeycomb lattice distorts, forming four shorter bonds and 2 longer bonds on each Ir hexagon.  (d) Above 3 GPa, the honeycomb lattice buckles and begins to dimerize, with each hexagon developing 2 short bonds, 2 medium bonds, and 2 long bonds.}
\label{fig:xrd}
\end{figure*}

In order to elucidate the role of structure in these electronic changes, we performed x-ray powder diffraction measurements, as shown in Fig.~\ref{fig:xrd}.  These measurements reveal that $\alpha$-$\rm Li_2IrO_3$ undergoes a series of two structural distortions as a function of pressure.  The first of these distortions, which arises at $P \sim 0.1$~GPa, is characterized by a gradual elongation of the Ir honeycomb lattice.  At ambient pressure, $\alpha$-$\rm Li_2IrO_3$ displays an almost ideal, undistorted Ir honeycomb lattice \cite{OMalley2008, Gretarsson2013a, Freund2016}, with 6 equal Ir-Ir bond lengths of 2.98~\AA.  By 0.1~GPa, we find that this honeycomb lattice has begun to distort, forming 2 long bonds (3.08~\AA) and 4 short bonds (2.92~\AA) on each Ir hexagon.  Such a distortion is fully allowed under the {\it C2/m} space group reported for this compound at ambient pressure.

This initial distortion is followed by a much larger distortion, which takes place during a first order structural phase transition at 3 GPa.  This transition is evident from peak splitting in the observed diffraction patterns (Fig.~\ref{fig:xrd}(a)), a discontinuous jump in lattice parameters (see SM), and an extended phase coexistence region from $P \sim 3$ to 5~GPa.  Structural refinements indicate that this transition is associated with a distortion that lowers the crystal symmetry from monoclinic ({\it C2/m}) to triclinic ({\it P-1}).  This causes the honeycomb lattice to stretch and buckle, with each Ir hexagon developing 2 short bonds, 2 intermediate bonds, and 2 long bonds.  The length of the 2 short bonds in the triclinic structure is remarkably small (2.31~\AA), which strongly suggests the formation of Ir-Ir dimers at high pressures.  A similar case of structural dimerization has also been reported in the honeycomb lattice ruthenate $\rm Li_2RuO_3$ \cite{Miura2007, Johannes2008, Kimber2014}.  In $\rm Li_2RuO_3$, the Ru honeycomb lattice exhibits a strong tendency to form local dimers and covalent Ru-Ru bonds \cite{Kimber2014}, with the development of long-range dimer order occurring below T$_C$ $\sim$ 540 K \cite{Miura2007, Kimber2014}.  Interestingly, we note that the rapid drop in branching ratio in $\alpha$-$\rm Li_2IrO_3$ appears to coincide with the small initial distortion at 0.1 GPa, rather than the much more obvious transition associated with the structural dimerization at 3 GPa.

\begin{figure*}
\begin{center}
\includegraphics[angle=0,width=\textwidth]{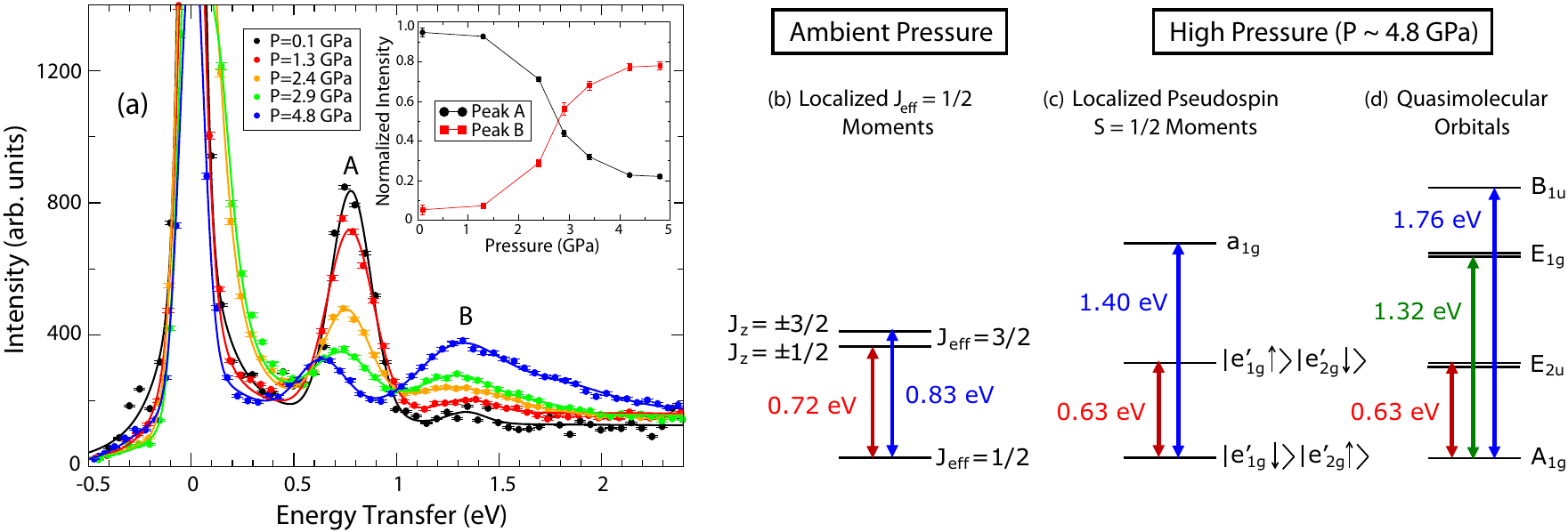}
\end{center}
\caption{Resonant inelastic x-ray scattering (RIXS) reveals the pressure dependence of the $d-d$ transitions and crystal electric field splittings in $\alpha$-$\rm Li_2IrO_3$.  (a) At ambient pressure, there is one strong peak (A) which corresponds to intra-$t_{2g}$ transitions between the $J_{eff}=3/2$ and $J_{eff}=1/2$ levels. Note that the small splitting of peak A is not observed in our low resolution setup. As the pressure increases, this peak drops rapidly in intensity, with a new peak (B) appearing at higher energies.  As shown in the inset, the combined spectral weight of these two features is approximately constant.  Quantitative analysis of the high pressure RIXS spectrum is consistent with a three-transition energy level scheme, which points towards the development of QMO-type physics.  The solid lines represent fits carried out using the procedure described in the Supplemental Material.  Potential $t_{2g}$ energy level schemes are provided for $\alpha$-$\rm Li_2IrO_3$ at (b) ambient pressure (localized $J_{eff}=1/2$), (c) P = 4.8 GPa (localized pseudospin S=1/2), and (d) P = 4.8 GPa (itinerant QMO picture).  These levels are illustrated in the hole representation, where a single $t_{2g}$ hole can be excited from the lowest energy level to the higher excited states.}
\label{fig:rixs}
\end{figure*}

The $J_{eff}=1/2$ relativistic electronic state arises from a very specific hierarchy of energy scales, set by crystal electric field, spin-orbit coupling, and electronic correlation effects.  These energy scales can be probed by resonant inelastic x-ray scattering, which is sensitive to $d-d$ transitions involving the Ir 5d valence levels (both within the $t_{2g}$ manifold and between the $t_{2g}$ and $e_g$ manifolds) \cite{Gretarsson2013a, JKim2012a, Liu2012, Hozoi2014, JKim2014}.  At ambient pressure, high resolution RIXS measurements on $\alpha$-$\rm Li_2IrO_3$ \cite{Gretarsson2013a} have shown that this material occupies a regime where octahedral crystal field splitting ($10Dq \sim 3.05$~eV) $\gg$ spin-orbit coupling ($3\lambda/2 \sim 0.78$~eV) $\gg$ trigonal crystal field splitting ($\Delta \sim 0.11$~meV).  These measurements indicate that the splitting of the lower $J_{eff}=3/2$ levels ($\Delta$) is small compared to the splitting between the $J_{eff}=3/2$ and $J_{eff}=1/2$ levels ($3\lambda/2$), providing some of the most compelling evidence in favor of the $J_{eff}=1/2$ description of this compound.

The pressure dependence of the RIXS spectra for $\alpha$-$\rm Li_2IrO_3$, obtained in a lower resolution configuration, is shown in Fig.~\ref{fig:rixs}.  This data indicates that the $d-d$ excitations are very sensitive to applied pressure, with significant changes in the distribution of spectral weight associated with transitions between the Ir $t_{2g}$ levels.  In particular, the strong energy loss peak at $\hbar \omega=E_i-E_f= 0.78$ eV (associated with transitions between the $J_{eff}=3/2$ and $J_{eff}=1/2$ levels) gradually decreases in intensity, while a new inelastic peak develops at $\hbar \omega \sim1.40$~eV.  The total spectral weight of these two features is approximately constant as a function of pressure (as shown in the inset of Fig.~\ref{fig:rixs}(a)), implying that spectral weight transfers from the low energy peak to the high energy peak, presumably due to a reorganization of the $t_{2g}$ energy levels.  The most obvious consequence of this new energy level scheme is that the trigonal crystal field splitting becomes larger than the spin-orbit coupling, confirming that the $J_{eff}=1/2$ model is no longer a valid description for this system.  We note that the peak at 1.40~eV first appears at the lowest applied pressure, but that the largest change in spectral weight coincides with the structural transition at $P \sim 3$~GPa.

A comparison of $t_{2g}$ energy level schemes corresponding to the localized $J_{eff}=1/2$, localized pseudospin S=1/2, and itinerant QMO models is provided in Fig. 3 (b,c,d).  Under moderate pressure (0.2 $<$ P $<$ 2.0 GPa), the RIXS spectra can be fit equally well using either the itinerant QMO model or the pseudospin S=1/2 model, a localized electron picture which applies in the limit of large crystal electric field ($\Delta > 3\lambda/2$) \cite{Subhro2012}.  However, the pressure-induced peak at 1.40~eV develops an increasingly asymmetric lineshape at higher pressure, and above $\sim$2 GPa it cannot be accurately fit using a single symmetric lineshape (see SM for further details).  The quality of fit is significantly improved by introducing a third inelastic peak at slightly higher energies ($\hbar \omega \sim 1.60$~eV).  Such a three peak spectrum cannot be justified in the localized electron model, but it is a distinguishing feature of the itinerant QMO model.

In the QMO model originally proposed for $\rm Na_2IrO_3$ by Mazin et al \cite{Mazin2012}, each Ir hexagon forms a series of six QMOs which are organized into four distinct energy levels as shown in Fig.~\ref{fig:rixs}.  These orbitals are occupied by 5 Ir valence electrons, giving rise to three possible d-d transitions within the $t_{2g}$ manifold.  Furthermore, the QMO theory predicts that the energies of these four levels are based on only two independent parameters: the nearest neighbor (NN) and next nearest neighbor (NNN) O-assisted hopping terms, $t_1^\prime$ and $t_2^\prime$.  The high pressure RIXS spectra can be fit to a model based on this QMO energy level scheme with remarkably good agreement.  The experimental values of the hopping parameters extracted from these fits are $t_1^\prime=0.27$~eV and $t_2^\prime=0.15$~eV at 2.4 GPa (monoclinic phase) and $t_1^\prime=0.33$~eV and $t_2^\prime=0.11$~eV at 4.8 GPa (triclinic phase).  These values can be compared to the theoretical estimates of $t_1^\prime=0.27$~eV and $t_2^\prime= 0.075$~eV predicted for $\rm Na_2IrO_3$ at ambient pressure \cite{Mazin2012}.

\begin{figure}
\begin{center}
\includegraphics[angle=0,width=3.0in]{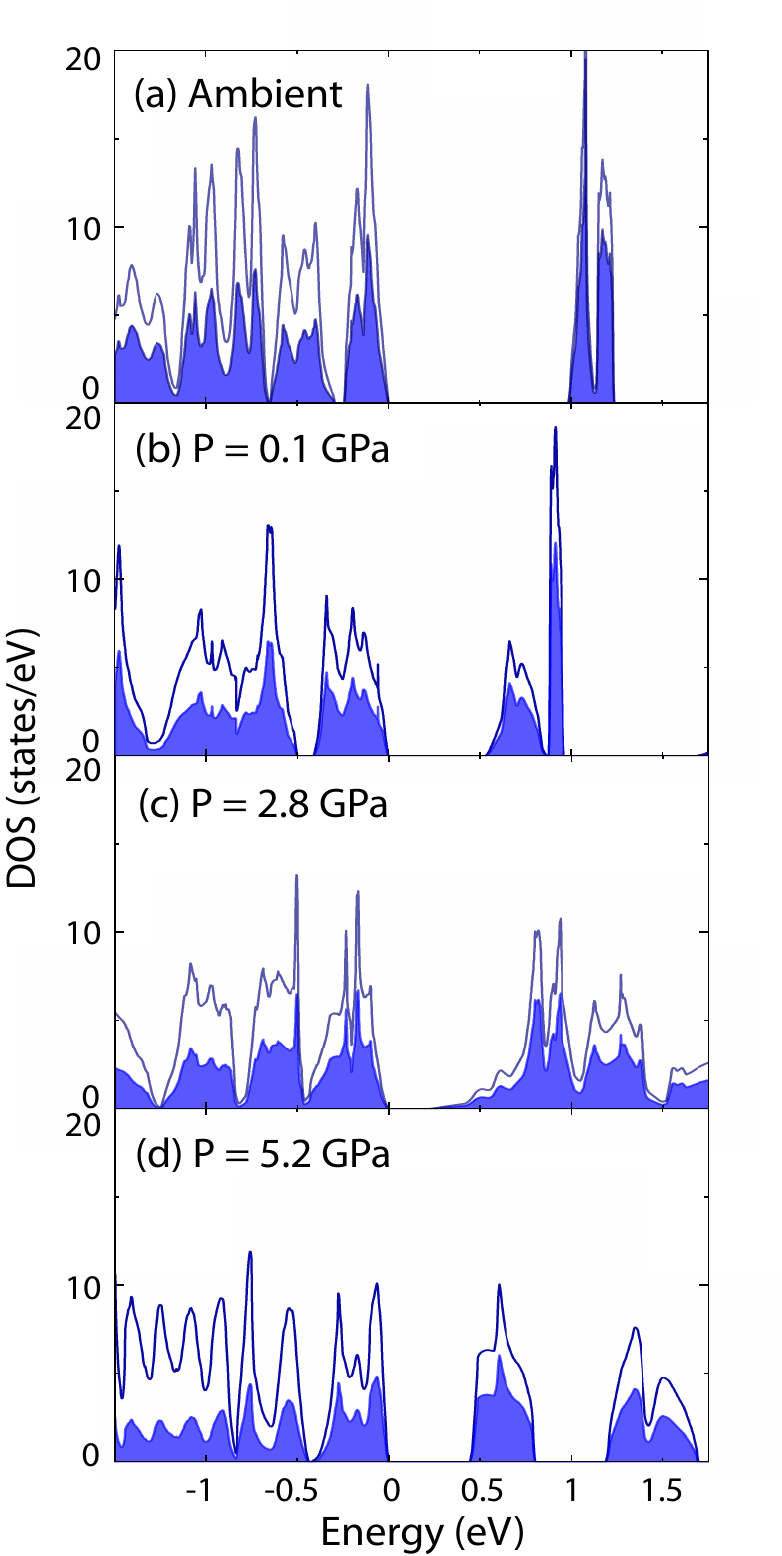}
\end{center}
\caption{The pressure dependence of the density of states (DOS) in $\alpha$-$\rm Li_2IrO_3$ can be investigated with density functional theory calculations (GGA + SOC + U, with Hubbard U = 2.0 eV and Hund's J = 0.5 eV) performed using the experimental crystal structures determined at (a) ambient, (b) P = 0.1 GPa, (c) P = 2.8 GPa, and (d) P = 5.2 GPa.  The solid curve represents the full DOS, while the shaded area represents the contribution due to the Ir electrons.  These calculations predict that $\alpha$-$\rm Li_2IrO_3$ should evolve from a $J_{eff}=1/2$ spin-orbital Mott insulator (ambient) towards a QMO insulator (P = 2.8 GPa) with increasing pressure.  This QMO state is characterized by a four peak DOS, with 3 peaks below the Fermi energy and 1 peak above.  In the dimerized triclinic phase (P = 5.2 GPa), the overlap between these peaks in the DOS increases significantly, indicating a potential breakdown of the QMO state.
}
\label{fig:dft}
\end{figure}

Additional support for the emergence of QMO-based physics is provided by the calculated density of states, as shown in Figure~\ref{fig:dft}. At ambient pressure (Fig.~\ref{fig:dft}(a)), the density of states resembles that of a localized $J_{eff}=1/2$ spin-orbital Mott insulator \cite{Gretarsson2013a}.  As the pressure increases towards 2.8 GPa (Figs. 4(b) and (c)), a series of four well-defined peaks develop in the vicinity of the Fermi level, with three peaks below $E_f$ and one peak above. This density of states is indicative of an itinerant QMO insulator with a gap of $\sim$ 0.2 eV.  The development of a QMO state is consistent with previous DFT calculations by Foyevtsova et al \cite{Mazin2013}, which suggest that moderate structural distortions act to enhance effective intrahexagon hopping parameters, while reducing the interhexagon hopping parameters.  As a result, the overall effect of the primary distortions that occur in the honeycomb iridate crystal structure - orthorhombic distortions, trigonal distortions, and rotations of the IrO$_6$ octahedra - is believed to enhance the QMO character of these materials. In recent theoretical work by Kim et al \cite{BHKim2016}, it has been argued that it should be possible to tune the honeycomb iridates between the localized $J_{eff}=1/2$ and itinerant QMO regimes by varying the energy scale associated with either the spin-orbit coupling or the electronic correlations.  Since hydrostatic pressure is often used to tune electronic correlations, it is reasonable to assume that a QMO state is revealed above 0.1 GPa as the strength of these correlations is reduced by applied pressure.

It should be noted that this QMO state may not survive into the heavily distorted triclinic phase above 3 GPa without some form of modification.  The calculated DOS in the dimerized phase (Fig. 4(d)) appears to have lost much of its QMO character, as the overlap between $t_{2g}$ bands below $E_f$ becomes significantly larger.  The detailed properties of this dimerized phase are a subject that still requires further investigation, however our experimental data clearly shows that this state remains insulating, and that it displays a RIXS spectrum that can still be well-described by a three peak, QMO-like energy level scheme.  It is intriguing to note that the energy level scheme and calculated DOS for dimerized $\rm Li_2RuO_3$ \cite{Johannes2008} also bears strong similarities to the QMO model.

Figure 5 presents a comparison of pressure scales identified by our three primary experimental techniques.  The pressure scales identified by XAS are illustrated by the pressure dependence of the Ir L$_2$ and L$_3$-edge white line intensities (Fig. 5(a)), the pressure scales identified by XRD are illustrated by the evolution of the Ir-Ir bond lengths (Fig. 5(b)), and the pressure scales identified by RIXS are illustrated by the evolution of the inelastic peak positions which correspond to energies of the intra-$t_{2g}$ transitions (Fig. 5(c)).  Taken in combination, these measurements point towards four distinct regimes:

\noindent (1) P $\lesssim$ 0.1 GPa - characterized by a high branching ratio, undistorted honeycomb lattice, and 2 peak RIXS spectrum.  These properties are consistent with a localized relativistic $J_{eff}$ = 1/2 ground state, as has generally been assumed for $\alpha$-$\rm Li_2IrO_3$ under ambient pressure conditions.

\noindent (2) 0.1 GPa $\lesssim$ P $\lesssim$ 2 GPa - characterized by a low branching ratio, slightly distorted honeycomb lattice, and 2 peak RIXS spectrum.  The drop in branching ratio implies a breakdown of the relativistic $J_{eff}$ = 1/2 ground state, and the energy level scheme can be explained in terms of either a localized pseudospin S = 1/2 model \cite{Subhro2012} or an itinerant QMO model \cite{Mazin2012}.

\noindent (3) 2 GPa $\lesssim$ P $\lesssim$ 3 GPa - characterized by a low branching ratio, slightly distorted honeycomb lattice, and 3 peak RIXS spectrum.  The evolution of the RIXS spectrum can no longer be explained in terms of a purely localized model, and is most naturally attributed to the development of an itinerant QMO ground state.

\noindent (4) P $\gtrsim$ 3 GPa - characterized by a low branching ratio, highly distorted honeycomb lattice, and 3 peak RIXS spectrum.  The large distortion of the crystal structure implies the development of a dimerized ground state.  The energy level scheme is still consistent with an itinerant QMO ground state.  However, it also displays strong similarities to the dimerized molecular orbital state of $\rm Li_2RuO_3$ \cite{Johannes2008}.

\begin{figure}
\begin{center}
\includegraphics[angle=0,width=3.0in]{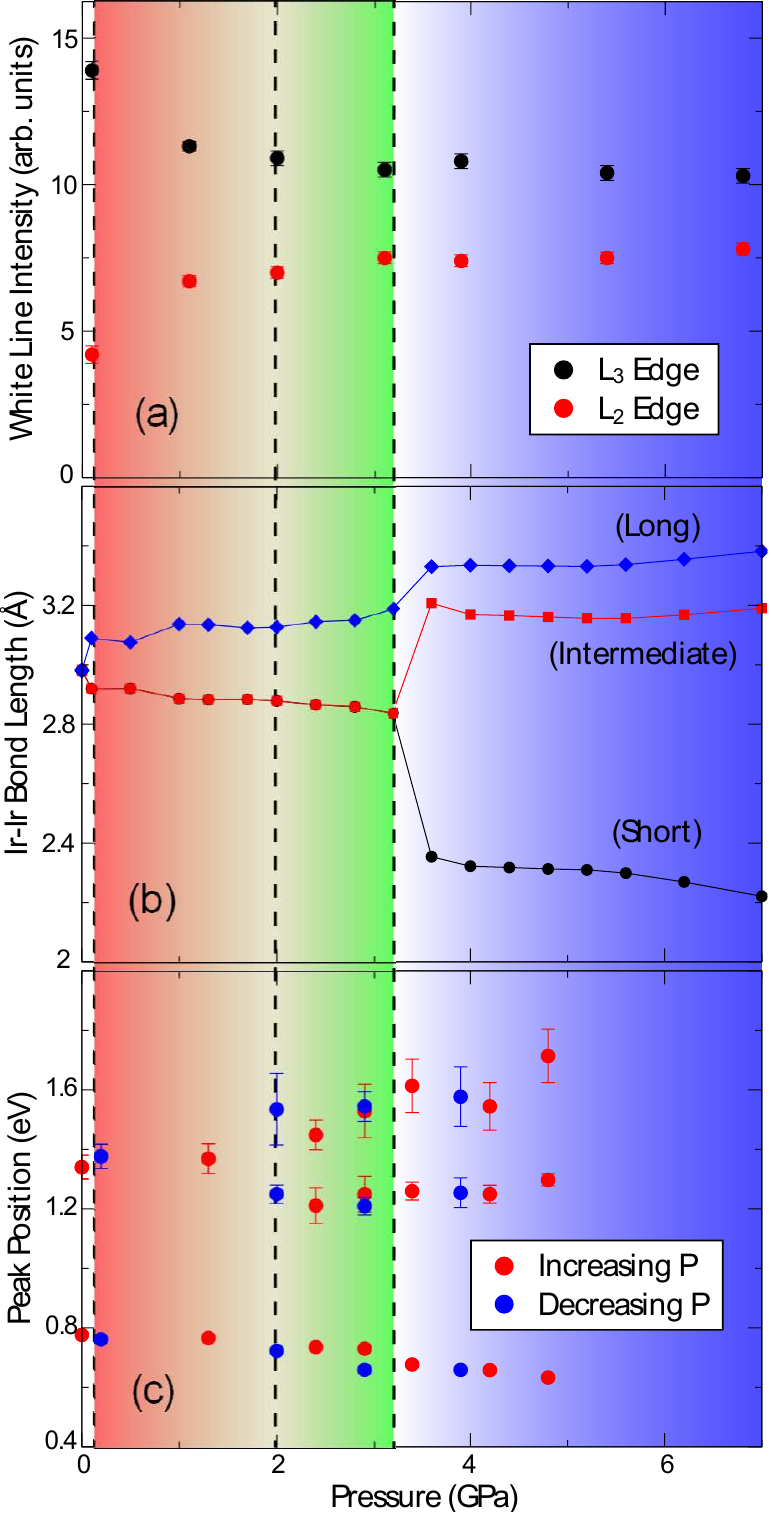}
\end{center}
\caption{A comparison of major pressure scales in $\alpha$-$\rm Li_2IrO_3$ identified by XAS, XRD, and RIXS.  (a) Pressure dependence of the Ir L$_2$ and L$_3$ edge white line intensity obtained from XAS.  (b) Pressure dependence of the Ir-Ir bond lengths determined from XRD.  (c) Pressure dependence of the inelastic peak positions (or intra-$t_{2g}$ transition energies) determined from RIXS.  The experimental data highlights four different regimes: a localized $J_{eff}$ = 1/2 state (unshaded), a localized pseudospin S=1/2 or itinerant QMO state (shaded red), an itinerant QMO state (shaded green), and an itinerant QMO or dimerized molecular orbital state (shaded blue).}\label{fig5}
\end{figure}

In summary, we present compelling experimental evidence of a pressure-driven collapse of the localized $J_{eff}=1/2$ relativistic electronic ground state in the honeycomb lattice iridate $\alpha$-$\rm Li_2IrO_3$.  Under the application of modest hydrostatic pressure, our complimentary x-ray diffraction and spectroscopy data clearly reveals a structural distortion which is accompanied by an electronic crossover from a localized $J_{eff}=1/2$ state to an itinerant QMO state. Our results show that the $J_{eff}=1/2$ state in $\alpha$-$\rm Li_2IrO_3$ at ambient pressure is extremely fragile, since it can be destroyed by a remarkably small pressure of 0.1 GPa.  Such fragility of the relativistic ground state could have important implications for the understanding of pressure-driven magnetic transitions in other Kitaev magnets.

{\it Note added in proof:} After the original submission of this article, high pressure x-ray diffraction measurements were reported on single crystal $\alpha$-$\rm Li_2IrO_3$ by V. Hermann {\it et al} \cite{Hermann2018}.  These measurements confirm the presence of a high pressure structural phase transition and structural dimerization at $P_{C}$ $\sim$ 3.8 GPa.

\section*{Acknowledgements}
Work at the University of Toronto was supported by the Natural Science and Engineering Research Council (NSERC) of Canada through the Collaborative Research and Training Experience (CREATE) program (432242-2013) and a Discovery Grant (RGPIN-2014-06071).  Research described in this paper was performed at the Canadian Light Source, which is supported by the Canada Foundation for Innovation, the Natural Sciences and Engineering Research Council of Canada, the University of Saskatchewan, the Government of Saskatchewan, Western Economic Diversification Canada, the National Research Council Canada, and the Canadian Institutes of Health Research.  In addition, this research used resources of the Advanced Photon Source, a U.S. Department of Energy (DOE) Office of Science User Facility operated for the DOE Office of Science by Argonne National Laboratory under Contract No. DE-AC02-06CH11357.


\begin{thebibliography}{31}
\expandafter\ifx\csname natexlab\endcsname\relax\def\natexlab#1{#1}\fi
\expandafter\ifx\csname bibnamefont\endcsname\relax
  \def\bibnamefont#1{#1}\fi
\expandafter\ifx\csname bibfnamefont\endcsname\relax
  \def\bibfnamefont#1{#1}\fi
\expandafter\ifx\csname citenamefont\endcsname\relax
  \def\citenamefont#1{#1}\fi
\expandafter\ifx\csname url\endcsname\relax
  \def\url#1{\texttt{#1}}\fi
\expandafter\ifx\csname urlprefix\endcsname\relax\def\urlprefix{URL }\fi
\providecommand{\bibinfo}[2]{#2}
\providecommand{\eprint}[2][]{\url{#2}}

\bibitem[{\citenamefont{Witczak-Krempa
  et~al.}(2014)\citenamefont{Witczak-Krempa, Chen, Kim, and
  Balents}}]{Witczak2014}
\bibinfo{author}{\bibfnamefont{W.}~\bibnamefont{Witczak-Krempa}},
  \bibinfo{author}{\bibfnamefont{G.}~\bibnamefont{Chen}},
  \bibinfo{author}{\bibfnamefont{Y.~B.} \bibnamefont{Kim}}, \bibnamefont{and}
  \bibinfo{author}{\bibfnamefont{L.}~\bibnamefont{Balents}},
  \bibinfo{journal}{Annual Review of Condensed Matter Physics}
  \textbf{\bibinfo{volume}{5}}, \bibinfo{pages}{57} (\bibinfo{year}{2014}).

\bibitem[{\citenamefont{Rau et~al.}(2016)\citenamefont{Rau, Lee, and
  Kee}}]{Rau2016}
\bibinfo{author}{\bibfnamefont{J.~G.} \bibnamefont{Rau}},
  \bibinfo{author}{\bibfnamefont{E.~K.-H.} \bibnamefont{Lee}},
  \bibnamefont{and} \bibinfo{author}{\bibfnamefont{H.-Y.} \bibnamefont{Kee}},
  \bibinfo{journal}{Annu. Rev. Condens. Matter Phys.}
  \textbf{\bibinfo{volume}{7}}, \bibinfo{pages}{195} (\bibinfo{year}{2016}).

\bibitem[{\citenamefont{Kim et~al.}(2008)\citenamefont{Kim, Jin, Moon, Kim,
  Park, Leem, Yu, Noh, Kim, Oh et~al.}}]{BJKim2008}
\bibinfo{author}{\bibfnamefont{B.~J.} \bibnamefont{Kim}},
  \bibinfo{author}{\bibfnamefont{H.}~\bibnamefont{Jin}},
  \bibinfo{author}{\bibfnamefont{S.~J.} \bibnamefont{Moon}},
  \bibinfo{author}{\bibfnamefont{J.-Y.} \bibnamefont{Kim}},
  \bibinfo{author}{\bibfnamefont{B.-G.} \bibnamefont{Park}},
  \bibinfo{author}{\bibfnamefont{C.~S.} \bibnamefont{Leem}},
  \bibinfo{author}{\bibfnamefont{J.}~\bibnamefont{Yu}},
  \bibinfo{author}{\bibfnamefont{T.~W.} \bibnamefont{Noh}},
  \bibinfo{author}{\bibfnamefont{C.}~\bibnamefont{Kim}},
  \bibinfo{author}{\bibfnamefont{S.-J.} \bibnamefont{Oh}},
  \bibnamefont{et~al.}, \bibinfo{journal}{Phys. Rev. Lett.}
  \textbf{\bibinfo{volume}{101}}, \bibinfo{pages}{076402}
  (\bibinfo{year}{2008}).

\bibitem[{\citenamefont{Kim et~al.}(2009)\citenamefont{Kim, Ohsumi, Komesu,
  Sakai, Morita, Takagi, and Arima}}]{BJKim2009}
\bibinfo{author}{\bibfnamefont{B.~J.} \bibnamefont{Kim}},
  \bibinfo{author}{\bibfnamefont{H.}~\bibnamefont{Ohsumi}},
  \bibinfo{author}{\bibfnamefont{T.}~\bibnamefont{Komesu}},
  \bibinfo{author}{\bibfnamefont{S.}~\bibnamefont{Sakai}},
  \bibinfo{author}{\bibfnamefont{T.}~\bibnamefont{Morita}},
  \bibinfo{author}{\bibfnamefont{H.}~\bibnamefont{Takagi}}, \bibnamefont{and}
  \bibinfo{author}{\bibfnamefont{T.}~\bibnamefont{Arima}},
  \bibinfo{journal}{Science} \textbf{\bibinfo{volume}{323}},
  \bibinfo{pages}{1329} (\bibinfo{year}{2009}).

\bibitem[{\citenamefont{Jackeli and Khaliullin}(2009)}]{Jackeli2009}
\bibinfo{author}{\bibfnamefont{G.}~\bibnamefont{Jackeli}} \bibnamefont{and}
  \bibinfo{author}{\bibfnamefont{G.}~\bibnamefont{Khaliullin}},
  \bibinfo{journal}{Phys. Rev. Lett.} \textbf{\bibinfo{volume}{102}},
  \bibinfo{pages}{017205} (\bibinfo{year}{2009}).

\bibitem[{\citenamefont{Chaloupka et~al.}(2010)\citenamefont{Chaloupka,
  Jackeli, and Khaliullin}}]{Chaloupka2010}
\bibinfo{author}{\bibfnamefont{J.}~\bibnamefont{Chaloupka}},
  \bibinfo{author}{\bibfnamefont{G.}~\bibnamefont{Jackeli}}, \bibnamefont{and}
  \bibinfo{author}{\bibfnamefont{G.}~\bibnamefont{Khaliullin}},
  \bibinfo{journal}{Phys. Rev. Lett.} \textbf{\bibinfo{volume}{105}},
  \bibinfo{pages}{027204} (\bibinfo{year}{2010}).

\bibitem[{\citenamefont{Singh et~al.}(2012)\citenamefont{Singh, Manni, Reuther,
  Berlijn, Thomale, Ku, Trebst, and Gegenwart}}]{Singh2012}
\bibinfo{author}{\bibfnamefont{Y.}~\bibnamefont{Singh}},
  \bibinfo{author}{\bibfnamefont{S.}~\bibnamefont{Manni}},
  \bibinfo{author}{\bibfnamefont{J.}~\bibnamefont{Reuther}},
  \bibinfo{author}{\bibfnamefont{T.}~\bibnamefont{Berlijn}},
  \bibinfo{author}{\bibfnamefont{R.}~\bibnamefont{Thomale}},
  \bibinfo{author}{\bibfnamefont{W.}~\bibnamefont{Ku}},
  \bibinfo{author}{\bibfnamefont{S.}~\bibnamefont{Trebst}}, \bibnamefont{and}
  \bibinfo{author}{\bibfnamefont{P.}~\bibnamefont{Gegenwart}},
  \bibinfo{journal}{Phys. Rev. Lett.} \textbf{\bibinfo{volume}{108}},
  \bibinfo{pages}{127203} (\bibinfo{year}{2012}).

\bibitem[{\citenamefont{Kitaev}(2006)}]{Kitaev2006}
\bibinfo{author}{\bibfnamefont{A.}~\bibnamefont{Kitaev}},
  \bibinfo{journal}{Annals of Physics} \textbf{\bibinfo{volume}{321}},
  \bibinfo{pages}{2 } (\bibinfo{year}{2006}).

\bibitem[{\citenamefont{Mazin et~al.}(2012)\citenamefont{Mazin, Jeschke,
  Foyevtsova, Valent\'\i, and Khomskii}}]{Mazin2012}
\bibinfo{author}{\bibfnamefont{I.~I.} \bibnamefont{Mazin}},
  \bibinfo{author}{\bibfnamefont{H.~O.} \bibnamefont{Jeschke}},
  \bibinfo{author}{\bibfnamefont{K.}~\bibnamefont{Foyevtsova}},
  \bibinfo{author}{\bibfnamefont{R.}~\bibnamefont{Valent\'\i}},
  \bibnamefont{and} \bibinfo{author}{\bibfnamefont{D.~I.}
  \bibnamefont{Khomskii}}, \bibinfo{journal}{Phys. Rev. Lett.}
  \textbf{\bibinfo{volume}{109}}, \bibinfo{pages}{197201}
  (\bibinfo{year}{2012}).

\bibitem[{\citenamefont{Foyevtsova et~al.}(2013)\citenamefont{Foyevtsova,
  Jeschke, Mazin, Khomskii, and Valenti}}]{Mazin2013}
\bibinfo{author}{\bibfnamefont{K.}~\bibnamefont{Foyevtsova}},
  \bibinfo{author}{\bibfnamefont{H.~O.} \bibnamefont{Jeschke}},
  \bibinfo{author}{\bibfnamefont{I.~I.} \bibnamefont{Mazin}},
  \bibinfo{author}{\bibfnamefont{D.~I.} \bibnamefont{Khomskii}},
  \bibnamefont{and} \bibinfo{author}{\bibfnamefont{R.}~\bibnamefont{Valenti}},
  \bibinfo{journal}{Phys. Rev. B} \textbf{\bibinfo{volume}{88}},
  \bibinfo{pages}{035107} (\bibinfo{year}{2013}).

\bibitem[{\citenamefont{Haskel et~al.}(2012)\citenamefont{Haskel, Fabbris,
  Zhernenkov, Kong, Jin, Cao, and van Veenendaal}}]{Haskel2012}
\bibinfo{author}{\bibfnamefont{D.}~\bibnamefont{Haskel}},
  \bibinfo{author}{\bibfnamefont{G.}~\bibnamefont{Fabbris}},
  \bibinfo{author}{\bibfnamefont{M.}~\bibnamefont{Zhernenkov}},
  \bibinfo{author}{\bibfnamefont{P.~P.} \bibnamefont{Kong}},
  \bibinfo{author}{\bibfnamefont{C.~Q.} \bibnamefont{Jin}},
  \bibinfo{author}{\bibfnamefont{G.}~\bibnamefont{Cao}}, \bibnamefont{and}
  \bibinfo{author}{\bibfnamefont{M.}~\bibnamefont{van Veenendaal}},
  \bibinfo{journal}{Phys. Rev. Lett.} \textbf{\bibinfo{volume}{109}},
  \bibinfo{pages}{027204} (\bibinfo{year}{2012}).

\bibitem[{\citenamefont{Tafti et~al.}(2012)\citenamefont{Tafti, Ishikawa,
  McCollam, Nakatsuji, and Julian}}]{Tafti2012}
\bibinfo{author}{\bibfnamefont{F.~F.} \bibnamefont{Tafti}},
  \bibinfo{author}{\bibfnamefont{J.~J.} \bibnamefont{Ishikawa}},
  \bibinfo{author}{\bibfnamefont{A.}~\bibnamefont{McCollam}},
  \bibinfo{author}{\bibfnamefont{S.}~\bibnamefont{Nakatsuji}},
  \bibnamefont{and} \bibinfo{author}{\bibfnamefont{S.~R.}
  \bibnamefont{Julian}}, \bibinfo{journal}{Phys. Rev. B}
  \textbf{\bibinfo{volume}{85}}, \bibinfo{pages}{205104}
  (\bibinfo{year}{2012}).

\bibitem[{\citenamefont{Clancy et~al.}(2016)\citenamefont{Clancy, Gretarsson,
  Lee, Tian, Kim, Upton, Casa, Gog, Islam, Jeon et~al.}}]{Clancy2016}
\bibinfo{author}{\bibfnamefont{J.~P.} \bibnamefont{Clancy}},
  \bibinfo{author}{\bibfnamefont{H.}~\bibnamefont{Gretarsson}},
  \bibinfo{author}{\bibfnamefont{E.~K.~H.} \bibnamefont{Lee}},
  \bibinfo{author}{\bibfnamefont{D.}~\bibnamefont{Tian}},
  \bibinfo{author}{\bibfnamefont{J.}~\bibnamefont{Kim}},
  \bibinfo{author}{\bibfnamefont{M.~H.} \bibnamefont{Upton}},
  \bibinfo{author}{\bibfnamefont{D.}~\bibnamefont{Casa}},
  \bibinfo{author}{\bibfnamefont{T.}~\bibnamefont{Gog}},
  \bibinfo{author}{\bibfnamefont{Z.}~\bibnamefont{Islam}},
  \bibinfo{author}{\bibfnamefont{B.-G.} \bibnamefont{Jeon}},
  \bibnamefont{et~al.}, \bibinfo{journal}{Phys. Rev. B}
  \textbf{\bibinfo{volume}{94}}, \bibinfo{pages}{024408}
  (\bibinfo{year}{2016}).

\bibitem[{\citenamefont{Laguna-Marco et~al.}(2010)\citenamefont{Laguna-Marco,
  Haskel, Souza-Neto, Lang, Krishnamurthy, Chikara, Cao, and van
  Veenendaal}}]{Laguna2010}
\bibinfo{author}{\bibfnamefont{M.~A.} \bibnamefont{Laguna-Marco}},
  \bibinfo{author}{\bibfnamefont{D.}~\bibnamefont{Haskel}},
  \bibinfo{author}{\bibfnamefont{N.}~\bibnamefont{Souza-Neto}},
  \bibinfo{author}{\bibfnamefont{J.~C.} \bibnamefont{Lang}},
  \bibinfo{author}{\bibfnamefont{V.~V.} \bibnamefont{Krishnamurthy}},
  \bibinfo{author}{\bibfnamefont{S.}~\bibnamefont{Chikara}},
  \bibinfo{author}{\bibfnamefont{G.}~\bibnamefont{Cao}}, \bibnamefont{and}
  \bibinfo{author}{\bibfnamefont{M.}~\bibnamefont{van Veenendaal}},
  \bibinfo{journal}{Phys. Rev. Lett.} \textbf{\bibinfo{volume}{105}},
  \bibinfo{pages}{216407} (\bibinfo{year}{2010}).

\bibitem[{\citenamefont{Clancy et~al.}(2012)\citenamefont{Clancy, Chen, Kim,
  Chen, Plumb, Jeon, Noh, and Kim}}]{Clancy2012}
\bibinfo{author}{\bibfnamefont{J.~P.} \bibnamefont{Clancy}},
  \bibinfo{author}{\bibfnamefont{N.}~\bibnamefont{Chen}},
  \bibinfo{author}{\bibfnamefont{C.~Y.} \bibnamefont{Kim}},
  \bibinfo{author}{\bibfnamefont{W.~F.} \bibnamefont{Chen}},
  \bibinfo{author}{\bibfnamefont{K.~W.} \bibnamefont{Plumb}},
  \bibinfo{author}{\bibfnamefont{B.~C.} \bibnamefont{Jeon}},
  \bibinfo{author}{\bibfnamefont{T.~W.} \bibnamefont{Noh}}, \bibnamefont{and}
  \bibinfo{author}{\bibfnamefont{Y.-J.} \bibnamefont{Kim}},
  \bibinfo{journal}{Phys. Rev. B} \textbf{\bibinfo{volume}{86}},
  \bibinfo{pages}{195131} (\bibinfo{year}{2012}).

\bibitem[{\citenamefont{van~der Laan and Thole}(1988)}]{VanderLaan1988}
\bibinfo{author}{\bibfnamefont{G.}~\bibnamefont{van~der Laan}}
  \bibnamefont{and} \bibinfo{author}{\bibfnamefont{B.~T.} \bibnamefont{Thole}},
  \bibinfo{journal}{Phys. Rev. Lett.} \textbf{\bibinfo{volume}{60}},
  \bibinfo{pages}{1977} (\bibinfo{year}{1988}).

\bibitem[{\citenamefont{Thole and van~der Laan}(1988)}]{Thole1988}
\bibinfo{author}{\bibfnamefont{B.~T.} \bibnamefont{Thole}} \bibnamefont{and}
  \bibinfo{author}{\bibfnamefont{G.}~\bibnamefont{van~der Laan}},
  \bibinfo{journal}{Phys. Rev. B} \textbf{\bibinfo{volume}{38}},
  \bibinfo{pages}{3158} (\bibinfo{year}{1988}).

\bibitem[{\citenamefont{Qi et~al.}(1987)\citenamefont{Qi, Perez, Ansari, Lu,
  and Croft}}]{Qi1987}
\bibinfo{author}{\bibfnamefont{B.}~\bibnamefont{Qi}},
  \bibinfo{author}{\bibfnamefont{I.}~\bibnamefont{Perez}},
  \bibinfo{author}{\bibfnamefont{P.~H.} \bibnamefont{Ansari}},
  \bibinfo{author}{\bibfnamefont{F.}~\bibnamefont{Lu}}, \bibnamefont{and}
  \bibinfo{author}{\bibfnamefont{M.}~\bibnamefont{Croft}},
  \bibinfo{journal}{Phys. Rev. B} \textbf{\bibinfo{volume}{36}},
  \bibinfo{pages}{2972} (\bibinfo{year}{1987}).

\bibitem[{\citenamefont{Jeon et~al.}(1989)\citenamefont{Jeon, Qi, Lu, and
  Croft}}]{Jeon1989}
\bibinfo{author}{\bibfnamefont{Y.}~\bibnamefont{Jeon}},
  \bibinfo{author}{\bibfnamefont{B.}~\bibnamefont{Qi}},
  \bibinfo{author}{\bibfnamefont{F.}~\bibnamefont{Lu}}, \bibnamefont{and}
  \bibinfo{author}{\bibfnamefont{M.}~\bibnamefont{Croft}},
  \bibinfo{journal}{Phys. Rev. B} \textbf{\bibinfo{volume}{40}},
  \bibinfo{pages}{1538} (\bibinfo{year}{1989}).

\bibitem[{\citenamefont{O'Malley et~al.}(2008)\citenamefont{O'Malley, Verweij,
  and Woodward}}]{OMalley2008}
\bibinfo{author}{\bibfnamefont{M.~J.} \bibnamefont{O'Malley}},
  \bibinfo{author}{\bibfnamefont{H.}~\bibnamefont{Verweij}}, \bibnamefont{and}
  \bibinfo{author}{\bibfnamefont{P.~M.} \bibnamefont{Woodward}},
  \bibinfo{journal}{J. Solid State Chem.} \textbf{\bibinfo{volume}{181}},
  \bibinfo{pages}{1803} (\bibinfo{year}{2008}).

\bibitem[{\citenamefont{Gretarsson et~al.}(2013)\citenamefont{Gretarsson,
  Clancy, Liu, Hill, Bozin, Singh, Manni, Gegenwart, Kim, Said
  et~al.}}]{Gretarsson2013a}
\bibinfo{author}{\bibfnamefont{H.}~\bibnamefont{Gretarsson}},
  \bibinfo{author}{\bibfnamefont{J.~P.} \bibnamefont{Clancy}},
  \bibinfo{author}{\bibfnamefont{X.}~\bibnamefont{Liu}},
  \bibinfo{author}{\bibfnamefont{J.~P.} \bibnamefont{Hill}},
  \bibinfo{author}{\bibfnamefont{E.}~\bibnamefont{Bozin}},
  \bibinfo{author}{\bibfnamefont{Y.}~\bibnamefont{Singh}},
  \bibinfo{author}{\bibfnamefont{S.}~\bibnamefont{Manni}},
  \bibinfo{author}{\bibfnamefont{P.}~\bibnamefont{Gegenwart}},
  \bibinfo{author}{\bibfnamefont{J.}~\bibnamefont{Kim}},
  \bibinfo{author}{\bibfnamefont{A.~H.} \bibnamefont{Said}},
  \bibnamefont{et~al.}, \bibinfo{journal}{Phys. Rev. Lett.}
  \textbf{\bibinfo{volume}{110}}, \bibinfo{pages}{076402}
  (\bibinfo{year}{2013}).

\bibitem[{\citenamefont{Freund et~al.}(2016)\citenamefont{Freund, Williams,
  Johnson, Coldea, Gegenwart, and Jesche}}]{Freund2016}
\bibinfo{author}{\bibfnamefont{F.}~\bibnamefont{Freund}},
  \bibinfo{author}{\bibfnamefont{S.~C.} \bibnamefont{Williams}},
  \bibinfo{author}{\bibfnamefont{R.~D.} \bibnamefont{Johnson}},
  \bibinfo{author}{\bibfnamefont{R.}~\bibnamefont{Coldea}},
  \bibinfo{author}{\bibfnamefont{P.}~\bibnamefont{Gegenwart}},
  \bibnamefont{and} \bibinfo{author}{\bibfnamefont{A.}~\bibnamefont{Jesche}},
  \bibinfo{journal}{Scientific Reports} \textbf{\bibinfo{volume}{6}},
  \bibinfo{pages}{35362} (\bibinfo{year}{2016}).

\bibitem[{\citenamefont{Miura et~al.}(2007)\citenamefont{Miura, Yasui, Sato,
  Igawa, and Kakurai}}]{Miura2007}
\bibinfo{author}{\bibfnamefont{Y.}~\bibnamefont{Miura}},
  \bibinfo{author}{\bibfnamefont{Y.}~\bibnamefont{Yasui}},
  \bibinfo{author}{\bibfnamefont{M.}~\bibnamefont{Sato}},
  \bibinfo{author}{\bibfnamefont{N.}~\bibnamefont{Igawa}}, \bibnamefont{and}
  \bibinfo{author}{\bibfnamefont{K.}~\bibnamefont{Kakurai}},
  \bibinfo{journal}{J. Phys. Soc. Jpn.} \textbf{\bibinfo{volume}{76}},
  \bibinfo{pages}{033705} (\bibinfo{year}{2007}).

\bibitem[{\citenamefont{Johannes et~al.}(2008)\citenamefont{Johannes, Stux, and
  Swider-Lyons}}]{Johannes2008}
\bibinfo{author}{\bibfnamefont{M.~D.} \bibnamefont{Johannes}},
  \bibinfo{author}{\bibfnamefont{A.~M.} \bibnamefont{Stux}}, \bibnamefont{and}
  \bibinfo{author}{\bibfnamefont{K.~E.} \bibnamefont{Swider-Lyons}},
  \bibinfo{journal}{Phys. Rev. B} \textbf{\bibinfo{volume}{77}},
  \bibinfo{pages}{075124} (\bibinfo{year}{2008}).

\bibitem[{\citenamefont{Kimber et~al.}(2014)\citenamefont{Kimber, Mazin, Shen,
  Jeschke, Streltsov, Argyriou, Valenti, and Khomskii}}]{Kimber2014}
\bibinfo{author}{\bibfnamefont{S.~A.~J.} \bibnamefont{Kimber}},
  \bibinfo{author}{\bibfnamefont{I.~I.} \bibnamefont{Mazin}},
  \bibinfo{author}{\bibfnamefont{J.}~\bibnamefont{Shen}},
  \bibinfo{author}{\bibfnamefont{H.~O.} \bibnamefont{Jeschke}},
  \bibinfo{author}{\bibfnamefont{S.~V.} \bibnamefont{Streltsov}},
  \bibinfo{author}{\bibfnamefont{D.~N.} \bibnamefont{Argyriou}},
  \bibinfo{author}{\bibfnamefont{R.}~\bibnamefont{Valenti}}, \bibnamefont{and}
  \bibinfo{author}{\bibfnamefont{D.~I.} \bibnamefont{Khomskii}},
  \bibinfo{journal}{Phys. Rev. B} \textbf{\bibinfo{volume}{89}},
  \bibinfo{pages}{081408(R)} (\bibinfo{year}{2014}).

\bibitem[{\citenamefont{Kim et~al.}(2012)\citenamefont{Kim, Casa, Upton, Gog,
  Kim, Mitchell, van Veenendaal, Daghofer, van~den Brink, Khaliullin
  et~al.}}]{JKim2012a}
\bibinfo{author}{\bibfnamefont{J.}~\bibnamefont{Kim}},
  \bibinfo{author}{\bibfnamefont{D.}~\bibnamefont{Casa}},
  \bibinfo{author}{\bibfnamefont{M.~H.} \bibnamefont{Upton}},
  \bibinfo{author}{\bibfnamefont{T.}~\bibnamefont{Gog}},
  \bibinfo{author}{\bibfnamefont{Y.-J.} \bibnamefont{Kim}},
  \bibinfo{author}{\bibfnamefont{J.~F.} \bibnamefont{Mitchell}},
  \bibinfo{author}{\bibfnamefont{M.}~\bibnamefont{van Veenendaal}},
  \bibinfo{author}{\bibfnamefont{M.}~\bibnamefont{Daghofer}},
  \bibinfo{author}{\bibfnamefont{J.}~\bibnamefont{van~den Brink}},
  \bibinfo{author}{\bibfnamefont{G.}~\bibnamefont{Khaliullin}},
  \bibnamefont{et~al.}, \bibinfo{journal}{Phys. Rev. Lett.}
  \textbf{\bibinfo{volume}{108}}, \bibinfo{pages}{177003}
  (\bibinfo{year}{2012}).

\bibitem[{\citenamefont{Liu et~al.}(2012)\citenamefont{Liu, Katukuri, Hozoi,
  Yin, Dean, Upton, Kim, Casa, Said, Gog et~al.}}]{Liu2012}
\bibinfo{author}{\bibfnamefont{X.}~\bibnamefont{Liu}},
  \bibinfo{author}{\bibfnamefont{V.~M.} \bibnamefont{Katukuri}},
  \bibinfo{author}{\bibfnamefont{L.}~\bibnamefont{Hozoi}},
  \bibinfo{author}{\bibfnamefont{W.-G.} \bibnamefont{Yin}},
  \bibinfo{author}{\bibfnamefont{M.~P.~M.} \bibnamefont{Dean}},
  \bibinfo{author}{\bibfnamefont{M.~H.} \bibnamefont{Upton}},
  \bibinfo{author}{\bibfnamefont{J.}~\bibnamefont{Kim}},
  \bibinfo{author}{\bibfnamefont{D.}~\bibnamefont{Casa}},
  \bibinfo{author}{\bibfnamefont{A.}~\bibnamefont{Said}},
  \bibinfo{author}{\bibfnamefont{T.}~\bibnamefont{Gog}}, \bibnamefont{et~al.},
  \bibinfo{journal}{Phys. Rev. Lett.} \textbf{\bibinfo{volume}{109}},
  \bibinfo{pages}{157401} (\bibinfo{year}{2012}).

\bibitem[{\citenamefont{Hozoi et~al.}(2014)\citenamefont{Hozoi, Gretarsson,
  Clancy, Jeon, Lee, Kim, Yushankhai, Fulde, Casa, Gog et~al.}}]{Hozoi2014}
\bibinfo{author}{\bibfnamefont{L.}~\bibnamefont{Hozoi}},
  \bibinfo{author}{\bibfnamefont{H.}~\bibnamefont{Gretarsson}},
  \bibinfo{author}{\bibfnamefont{J.~P.} \bibnamefont{Clancy}},
  \bibinfo{author}{\bibfnamefont{B.-G.} \bibnamefont{Jeon}},
  \bibinfo{author}{\bibfnamefont{B.}~\bibnamefont{Lee}},
  \bibinfo{author}{\bibfnamefont{K.~H.} \bibnamefont{Kim}},
  \bibinfo{author}{\bibfnamefont{V.}~\bibnamefont{Yushankhai}},
  \bibinfo{author}{\bibfnamefont{P.}~\bibnamefont{Fulde}},
  \bibinfo{author}{\bibfnamefont{D.}~\bibnamefont{Casa}},
  \bibinfo{author}{\bibfnamefont{T.}~\bibnamefont{Gog}}, \bibnamefont{et~al.},
  \bibinfo{journal}{Phys. Rev. B} \textbf{\bibinfo{volume}{89}},
  \bibinfo{pages}{115111} (\bibinfo{year}{2014}).

\bibitem[{\citenamefont{Kim et~al.}(2014)\citenamefont{Kim, Daghofer, Said,
  Gog, van~den Brink, Khaliullin, and Kim}}]{JKim2014}
\bibinfo{author}{\bibfnamefont{J.}~\bibnamefont{Kim}},
  \bibinfo{author}{\bibfnamefont{M.}~\bibnamefont{Daghofer}},
  \bibinfo{author}{\bibfnamefont{A.~H.} \bibnamefont{Said}},
  \bibinfo{author}{\bibfnamefont{T.}~\bibnamefont{Gog}},
  \bibinfo{author}{\bibfnamefont{J.}~\bibnamefont{van~den Brink}},
  \bibinfo{author}{\bibfnamefont{G.}~\bibnamefont{Khaliullin}},
  \bibnamefont{and} \bibinfo{author}{\bibfnamefont{B.~J.} \bibnamefont{Kim}},
  \bibinfo{journal}{Nat. Comm.} \textbf{\bibinfo{volume}{5}},
  \bibinfo{pages}{4453} (\bibinfo{year}{2014}).

\bibitem[{\citenamefont{Bhattacharjee et~al.}(2012)\citenamefont{Bhattacharjee,
  Lee, and Kim}}]{Subhro2012}
\bibinfo{author}{\bibfnamefont{S.}~\bibnamefont{Bhattacharjee}},
  \bibinfo{author}{\bibfnamefont{S.-S.} \bibnamefont{Lee}}, \bibnamefont{and}
  \bibinfo{author}{\bibfnamefont{Y.~B.} \bibnamefont{Kim}},
  \bibinfo{journal}{New J. Phys.} \textbf{\bibinfo{volume}{14}},
  \bibinfo{pages}{073015} (\bibinfo{year}{2012}).

\bibitem[{\citenamefont{Kim et~al.}(2016)\citenamefont{Kim, Shirakawa, and
  Yunoki}}]{BHKim2016}
\bibinfo{author}{\bibfnamefont{B.~H.} \bibnamefont{Kim}},
  \bibinfo{author}{\bibfnamefont{T.}~\bibnamefont{Shirakawa}},
  \bibnamefont{and} \bibinfo{author}{\bibfnamefont{S.}~\bibnamefont{Yunoki}},
  \bibinfo{journal}{Phys. Rev. Lett.} \textbf{\bibinfo{volume}{117}},
  \bibinfo{pages}{187201} (\bibinfo{year}{2016}).
	
	\bibitem[{\citenamefont{Hermann et~al.}(2018)\citenamefont{Hermann, Altmeyer, Ebad-Allah, Freund, Jesche, Tsirlin, Hanfland, Gegenwart, Mazin, Khomskii, Valenti and 
	Kuntscher}}]{Hermann2018}
\bibinfo{author}{\bibfnamefont{V.} \bibnamefont{Hermann}},
  \bibinfo{author}{\bibfnamefont{M.}~\bibnamefont{Altmeyer}},
	\bibinfo{author}{\bibfnamefont{J.}~\bibnamefont{Ebad-Allah}},
	\bibinfo{author}{\bibfnamefont{F.}~\bibnamefont{Freund}},
	\bibinfo{author}{\bibfnamefont{A.}~\bibnamefont{Jesche}},
	\bibinfo{author}{\bibfnamefont{A. A.}~\bibnamefont{Tsirlin}},
	\bibinfo{author}{\bibfnamefont{M.}~\bibnamefont{Hanfland}},
	\bibinfo{author}{\bibfnamefont{P.}~\bibnamefont{Gegenwart}},
	\bibinfo{author}{\bibfnamefont{I. I.}~\bibnamefont{Mazin}},
	\bibinfo{author}{\bibfnamefont{D. I.}~\bibnamefont{Khomskii}},
	\bibinfo{author}{\bibfnamefont{R.}~\bibnamefont{Valenti}},
  \bibnamefont{and} \bibinfo{author}{\bibfnamefont{C. A.}~\bibnamefont{Kuntscher}},
  \bibinfo{journal}{Phys. Rev. B} \textbf{\bibinfo{volume}{97}},
  \bibinfo{pages}{020104(R)} (\bibinfo{year}{2018}).

\end{thebibliography}
\end{document}